\documentclass[figures,cite]{epl}
\usepackage{graphicx}
\usepackage{amsmath}
\usepackage[latin1]{inputenc}
\usepackage[T1]{fontenc}
\usepackage{ae,aecompl}

\title{Agreement dynamics on small-world networks}
\author{L. Dall'Asta\inst{1} \and A. Baronchelli\inst{2}
\and A. Barrat\inst{1} \and V. Loreto\inst{2}}
\institute{ 
  \inst{1}Laboratoire de
  Physique Th\'eorique (CNRS UMR8627) - B\^atiment 210,
  Universit\'e Paris-Sud, 91405 Orsay cedex, France\\
  \inst{2}Dipartimento di Fisica, Universit\`a ``La Sapienza'' and SMC-INFM,
P.le A. Moro 2, 00185 ROMA, (Italy)
}


\begin{document}
\maketitle
\begin{abstract}
In this paper we analyze the effect of a non-trivial topology on the
dynamics of the so-called Naming Game, a recently introduced model
which addresses the issue of how shared conventions emerge
spontaneously in a population of agents. We consider in particular the
small-world topology and study the convergence towards the global agreement
as a function of the population size $N$ as well as of the parameter
$p$ which sets the rate of rewiring leading to the small-world
network.  As long as $p \gg 1/N$ there exists a crossover time scaling as
$N/p^2$ which separates an early one-dimensional-like dynamics from a
late stage mean-field-like behavior. At the beginning of the process, the
local quasi one-dimensional topology induces a coarsening dynamics
which allows for a minimization of the cognitive effort (memory)
required to the agents. In the late stages, on the other hand, 
the mean-field like topology leads to a speed up of the convergence 
process with respect to the one-dimensional case. 
\end{abstract}

The recent past has witnessed an important development of the
activities of statistical physicists in the area of social sciences
(for a recent collection of papers see~\cite{Granada}). Indeed,
statistical physics is the natural field to study how 
global complex properties can emerge from purely local rules. 
Social interactions have thus been described and studied
by statistical physics
models and tools, in particular models
of opinion formation in which agents update their internal state, or
opinion, through an interaction with its neighbors. An interesting
point concerns whether and how a population of agents converge towards
a common and shared state (consensus) without external global
coordination~\cite{opinions}.  Early studies have mostly dealt with
agents either able to interact with all the other agents (mean-field
case), or sitting on the nodes of regular lattices. Such situations,
although not always realistic, have the advantage to be accessible to
the usual methods of statistical mechanics.

Even more recently however, the growing field of complex 
networks~\cite{bara,mdbook,psvbook} has
allowed to obtain a better knowledge of social
networks~\cite{granovetter}, and in particular to show that the
topology of the network on which agents interact is not regular. A
natural step has then been to consider various models
embedded on more realistic networks and to study the influence of
various complex topologies on the corresponding dynamical behavior.

In particular, social networks are typically ``small-worlds'' in
which, on the one hand, the average distance between two agents is
small~\cite{Milgram:1967}, growing only logarithmically with the
network's size, and, on the other hand, many triangles are present, unlike
totally random networks. In order to reconcile both properties,
Watts and Strogatz have introduced the now famous small-world network
model~\cite{Watts:1998} which allows to interpolate between regular
low-dimensional lattices and random networks, by introducing a certain
amount of random long-range connections into an initially regular
network.

Subsequently, a number of papers have focused on the influence of
these long-range ``short-cuts'' on the behavior of various models
defined on the network: from the Ising model~\cite{Barrat:2000} to the
spreading of epidemics~\cite{Moore:2000}, or the evolution of random
walks~\cite{Jasch:2001}. Dynamics of models inspired by social
sciences are no exception, such as the Voter
model~\cite{Castellano:2003,Vilone:2003} or Axelrod's model of culture
dissemination~\cite{Axelrod97,Kuperman:2005}.

In this letter, we consider the effect of a small-world topology on
the so-called Naming Game model, which was inspired by the 
field of semiotic dynamics, a new emerging area focusing on the 
development of shared communication systems (languages) among a population 
of agents. Such a process can indeed be considered as arising through 
self-organization out of local interactions. The language is then
seen as constantly reshaped by its users in order to maximize
communicative success and expressive power while minimizing the
cognitive effort~\cite{Steels1997,Kirby}. In addition, there are
recent developments in Information Technology in which new forms of
semiotic dynamics begin to appear. One example are social tagging
sites (such as del.icio.us or www.flickr.com), through which tens of
thousands of web users share information by tagging items like
pictures or web-sites and thus develop
"folksonomies"~\cite{huberman,ciro}. In this context, simplified
models or ``Language Games'' have been defined and studied in the
theoretical community~\cite{Steels1997,Kirby}. As for opinion
formation models, it is interesting to understand if a common state
for all agents can be reached and, in the positive answer case, how
the system converges towards such a state.

\section{The model}

The original model~\cite{Steels:1995} is related to an
artificial intelligence experiment called Talking
Heads~\cite{Steels1998}, in which embodied software agents observe a
set of objects through digital cameras, assign them randomly chosen
names and communicate these names to each other. Since different
agents can invent different names for the same object, the final
emergence of a common dictionary for all the agents is not granted
from the start. However, it turns out that such a consensus is in fact
experimentally reached. In order to try to capture the essential
relevant features of such a dynamics, Baronchelli et
al.~\cite{Baronchelli:2005} have proposed a minimal model of Naming
Game that reproduces the phenomenology of the experiments, despite the
agents of the model are far from the complicate software effectively
used as ``Talking Heads''. Such a model is however amenable to both
analytical and extensive numerical
treatment~\cite{Baronchelli:2005,NG1d}, allowing for a better
understanding of the mechanisms at work.

The model considers $N$ identical individuals (or agents) which
observe the same object and try to communicate its name one to the
other. Each agent is endowed with an internal inventory or memory in
which it can store an a priori unlimited number of different names or
opinions. Initially, each agent has an empty inventory. The
dynamics proceeds as follows: at each time step, two individuals are
chosen at random for a pairwise interaction (or
``communication''). One of these agents acts as the ``speaker'' and the
other one as the ``hearer''. If the speaker does not know a name for the
object (its inventory is empty), it invents a new name and records it. Else, if it already knows one or more synonyms (stored in the inventory), it
chooses one of them randomly. The invented or selected word is then transmitted
to the hearer. If the hearer already has this term in its memory, 
the interaction is a success, and both agents retain that term
as the right one, canceling all the other terms in their inventories;
otherwise, the interaction is a failure, and the new name is included
in the inventory of the hearer, without any cancellation.

The way in which agents may interact with each other is determined by
the topology of the underlying contact network. The mean-field case
corresponds to a fully connected network, in which all agents are in
mutual contact. In this case, studied in~\cite{Baronchelli:2005}, each
agent rarely interacts twice with the same partner, so that the system
initially accumulates a large number ($\mathcal{O}(N/2)$) of different 
names
(synonyms) for the object, invented by different agents (speakers) and
$\mathcal{O}(N^{3/2})$ total words in the whole population.  
Interestingly however, this profusion of different names leads in the end 
to an asymptotic absorbing state in which all the agents share the same
name.

As a second step towards the understanding of the model from a statistical
physics point of view, we have considered in~\cite{NG1d} the case of agents
sitting on the nodes of a regular lattice in dimension $d$. In this case, each
agent is connected to a finite number of neighbors ($2d$) so that it may
possess only a finite number of different words in its inventory at any given
time.  As a result, the total amount of memory used by the whole system 
grows as $N$ instead of $N^{3/2}$. Local consensus appears at very 
early stages of the evolution, since neighboring agents tend to share 
the same unique word. The dynamics then proceeds through the 
coarsening of such clusters of agents sharing a common name; the 
interfaces between clusters are composed by agents who still have more 
than one possible name, and diffuse
randomly. Because of this particular coarsening process, the average 
cluster size grows as $\sqrt{t/N}$, and the time to convergence
corresponds to the time needed for one cluster to reach the system size, i.e.
a time $N^{1+2/d}$ for $d\leq4$.  In one dimension in particular, the
convergence is thus dramatically slowed down from ${\cal O}(N^{3/2})$ to
${\cal O}(N^3)$.

In the following, we investigate the effect of long-range connections
which link agents that are far from each other on the regular lattice. We use
the small-world model of Watts and Strogatz:
starting from a one-dimensional lattice of $N$ sites, with periodic
boundary conditions (i.e. a ring), each vertex being connected to its
$2 m$ nearest neighbors, a stochastic rewiring procedure is applied.
The vertices are visited one after the other, and each link connecting
a vertex to one of its $m$ nearest neighbors in the clockwise sense
is left in place with probability $1-p$, and with probability $p$ is
reconnected to a randomly chosen other vertex. For $p=0$ the network
retains a purely one-dimensional topology, while the random network
structure is approached as $p$ goes to $1$. At small but finite $p$
($1/N \ll p \ll 1$), a small-world structure with short distances
between nodes, together with a large clustering, is obtained.

\section{Global picture: Expected behavior}

For $p=0$, it has been shown in~\cite{NG1d} that the dynamics proceeds
by a slow coarsening of clusters of agents sharing the same state or
word.  At small $p$, the short-cuts are typically far from each other,
with a typical distance $1/p$ between short-cuts so that the early
dynamics is not affected and proceeds as in dimension $1$. In
particular, at very short times many new words are invented since the
success rate is small.  After a time of order $N$, each agent has
played typically once, and therefore ${\cal O}(N)$ different words
have been invented: the number of different words reaches a peak
which scales as $N$.  Since the number of neighbors of each site is
bounded (the degree distribution decreases
exponentially~\cite{Barrat:2000}), each agent has access only to a
finite number of different words, so that the average memory per agent
used remains finite, as in finite dimensions and in contrast with the
mean-field case. The interaction of neighboring agents first leads to
the usual coarsening phenomena as long as the clusters are typically
one-dimensional, i.e. as long as the typical cluster size is smaller
than $1/p$. However, as the average cluster size reaches the typical
distance between two short-cuts $\sim 1/p$, a crossover phenomena is bound 
to take place; since the cluster size grows as $\sqrt{t/N}$~\cite{NG1d}, 
this corresponds to a crossover time $t_{cross} = {\cal O}(N/p^2)$. For
times much larger than this crossover, one expects that the dynamics
is dominated by the existence of short-cuts and enters a mean-field
like behavior. The convergence time is thus expected to scale as
$N^{3/2}$ and not as $N^3$. In order for this picture to be possible,
the crossover time $N/p^2$ needs to be much larger than $1$, and much
smaller than the consensus time for the one-dimensional case $N^3$;
these two conditions read $p \gg 1/N$, which is indeed the necessary
condition to obtain a small-world network.

It is therefore expected that the small-world topology allows to
combine advantages from both finite-dimensional lattices and
mean-field networks: on the one hand, only a finite memory per node is
needed, in opposition to the ${\cal O}(N^{1/2})$ in mean-field; on the
other hand the convergence time is expected to be much shorter than in
finite dimensions.

\section{Numerical study}

Various quantities of interest can be monitored in numerical studies
of the Naming Game model in order to verify and quantify the
qualitative expected picture. Among the most relevant ones are the average
number of words in the agents inventory, $N_w(t)$, which corresponds
to the average memory used, and the total number of distinct words in
the system, $N_d(t)$.

\begin{figure}
\centerline{ 
\includegraphics*[width=0.4\textwidth]{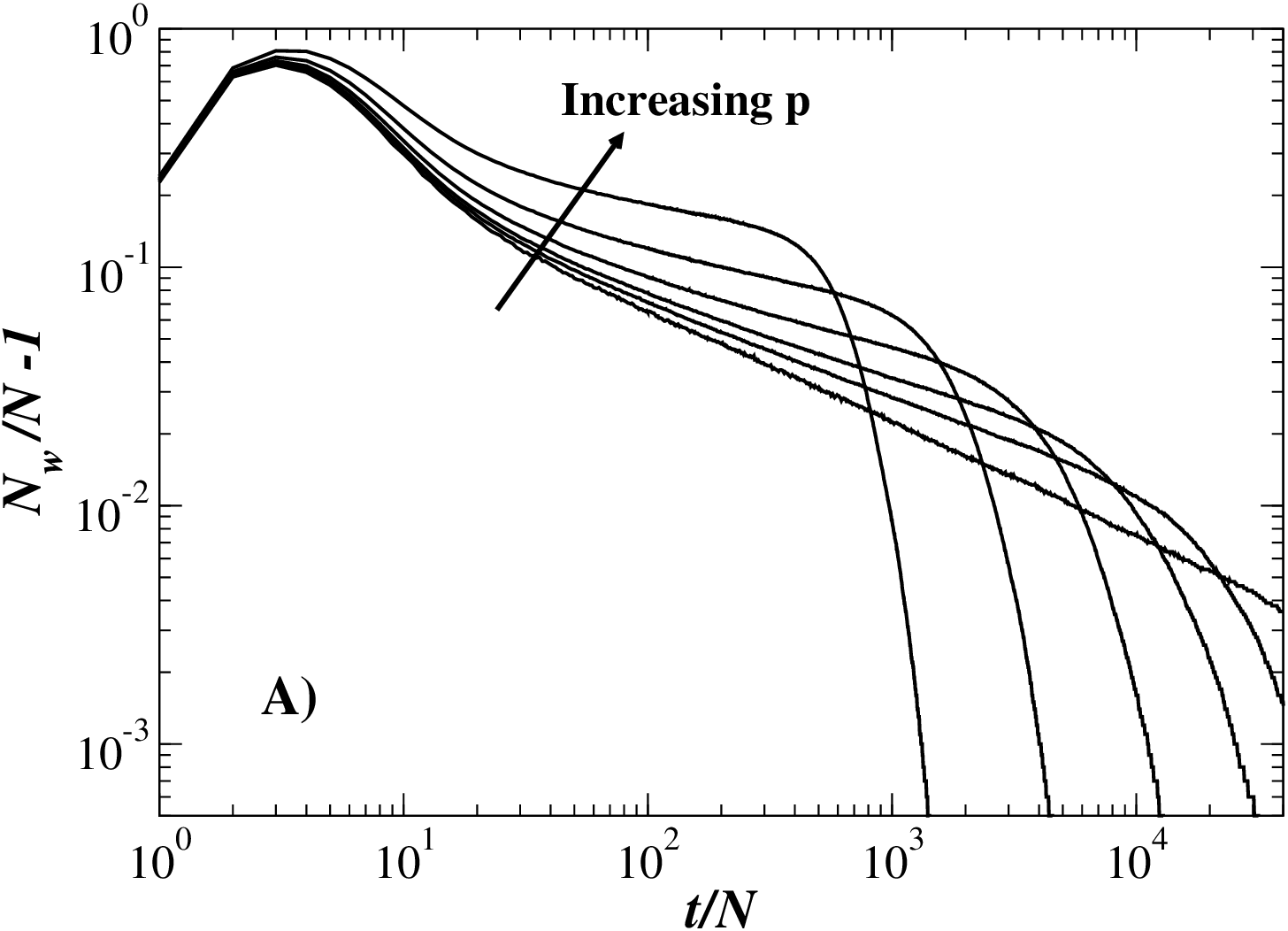}
\includegraphics*[width=0.4\textwidth]{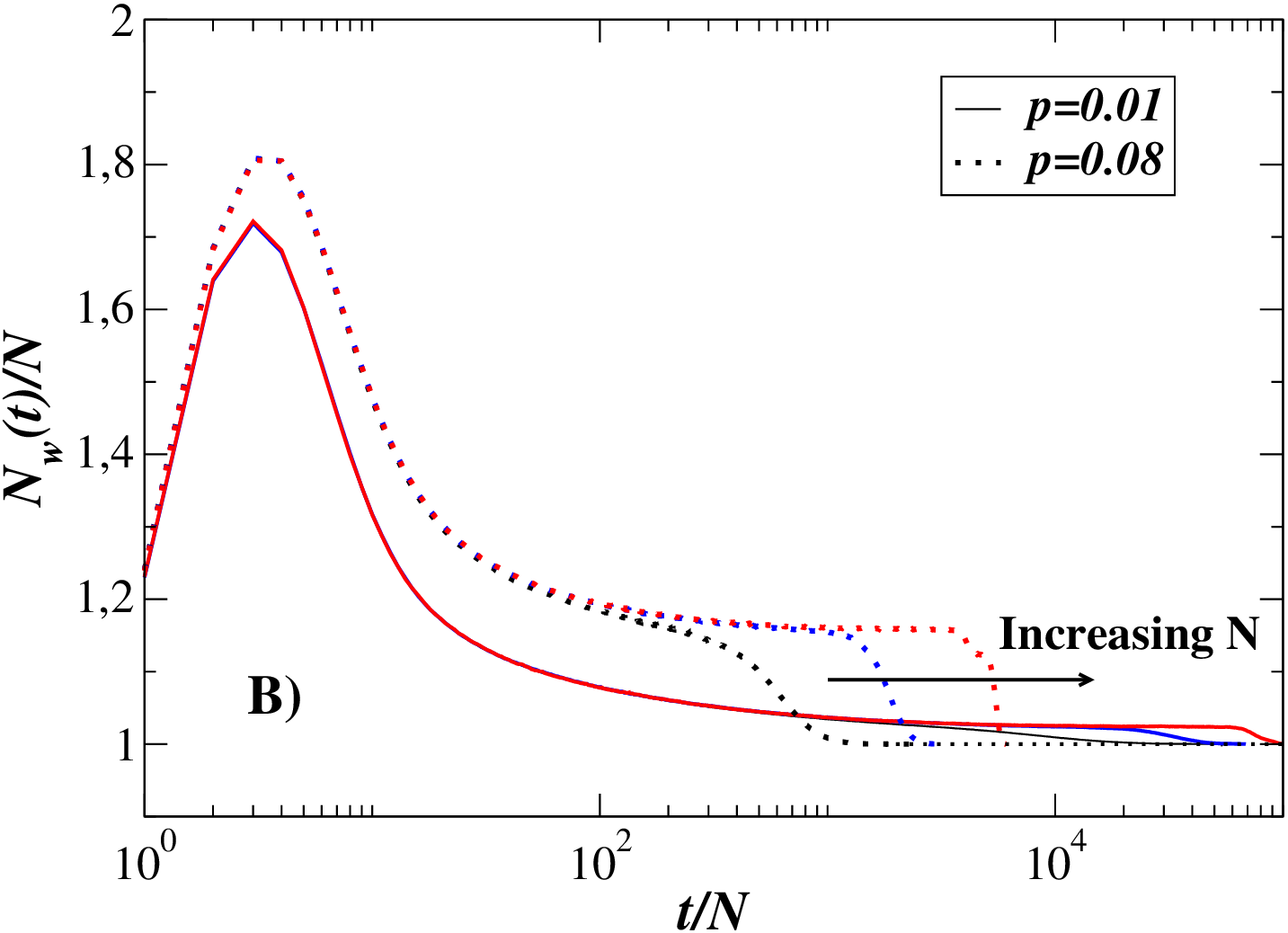}
}
\caption{A) Average number of words per agent in the system, $N_w/N$
as a function of the rescaled time $t/N$, for small-world networks
with $\langle k \rangle=8$ and $N=10^3$ nodes, for various values of
$p$. The curve for $p=0$ is shown for reference, as well as
$p=5. 10^{-3}$, $p=10^{-2}$, $p=2. 10^{-2}$, $p=4. 10^{-2}$,
$p=8. 10^{-2}$, from bottom to top on the left part of the curves.  
B) (Color Online) $N_w/N$ for $p=10^{-2}$ and $p=8. 10^{-2}$ and
increasing system sizes: $N=10^3$, $N=10^4$, $N=10^5$. Larger system sizes
yield larger plateau lengths.}
\label{fig:Nwk8}
\end{figure}

Figure~\ref{fig:Nwk8} displays the evolution of the average number of
words per agent as a function of time, for a small-world network with
average degree $\langle k \rangle=8$, and various values of the
rewiring probability $p$ and size $N$. While $N_w(t)$ in all cases
decays to $N$ (Fig.~\ref{fig:Nwk8}A), after an initial peak whose
height is proportional to $N$ (Fig.~\ref{fig:Nwk8}B), the way in which
this convergence is obtained depends on the parameters. At fixed $N$,
for $p=0$ a power-law behavior $N_w/N - 1 \propto 1/\sqrt{t}$ is
observed due to the one-dimensional coarsening process~\cite{NG1d}. As
soon as $p \gg 1/N$ however, deviations are observed and get stronger
as $p$ is increased: the decrease of $N_w$ is first slowed down after
the peak, but leads in the end to an exponential convergence. The
intermediate slowing down and the faster convergence are both enhanced
as $p$ increases. On the other hand, a system size increase at fixed
$p$ corresponds, as shown in Fig.~\ref{fig:Nwk8}B, to a slower
convergence {\em even on the rescaled time $t/N$}, with a longer and
longer plateau at almost constant average used memory.

\begin{figure}
\centerline{
\includegraphics*[width=0.4\textwidth]{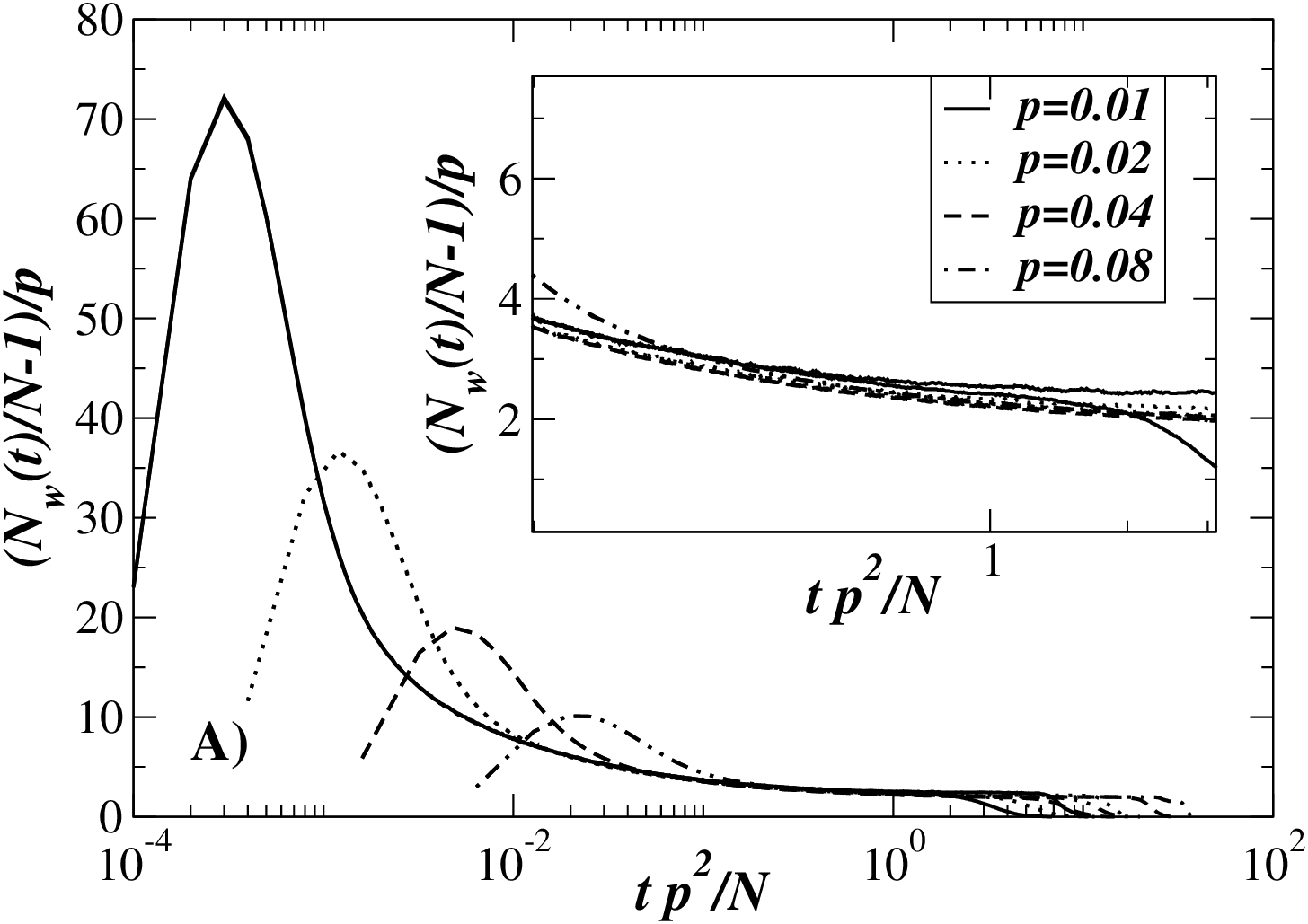}
\includegraphics*[width=0.4\textwidth]{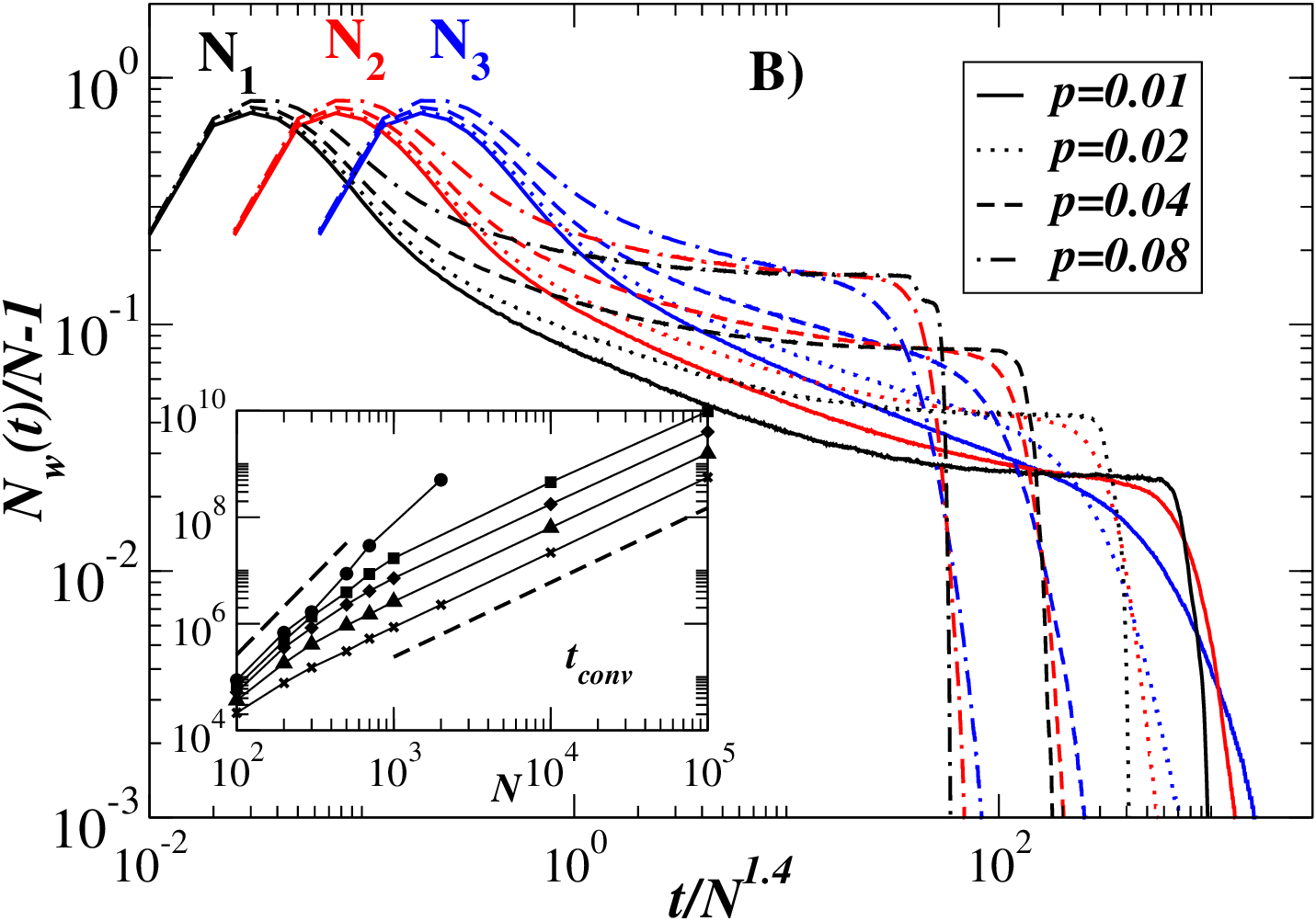}
}
\caption{A) Rescaled curves of the average number of words per agent
in the system, in order to show the collapse around the crossover time
$N/p^2$. For each value of $p$, two values of the system size
($N=10^4$ and $N=10^5$) are displayed. The curves for different sizes
are perfectly superimposed before the convergence.  
B) (Color online) Convergence at
large times, shown by the drop of $N_w/N-1$ to $0$: the time is
rescaled by $N^{1.4}$. For each $p$, three different sizes ($N_1=10^3$ for
the left peak, curves in black, 
$N_2=10^4$ for the middle peak, curves in blue, and $N_3=10^5$, right peak,
curves in red) 
are shown. On the $N^{1.4}$ scale, the
convergence becomes more and more abrupt as $N$ increases. The inset
displays the convergence time as a function of size for $p=0$
(bullets), $p=0.01$ (squares), $p=0.02$ (diamonds), $p=0.04$
(triangles), $p=0.08$ (crosses); the dashed lines are 
proportional to $N^3$ and $N^{1.4}$.}
\label{fig:Nwk8resc}
\end{figure}

As mentioned previously, a crossover phenomenon is expected when the
one-dimensional clusters reach sizes of order $1/p$, i.e. at a time of
order $N/p^2$. By definition, in the interior of each cluster, sites
have only one word in memory, while the sites with more than one word
are localized at the interfaces between clusters, whose number is then
of order $Np$.  The average excess memory per site (with respect to 
global consensus) is thus of order $p$, so that one
expects $N_w/N -1= p {\cal G}(tp^2/N)$. Figure~\ref{fig:Nwk8resc}A
indeed shows that the data of $(N_w/N -1)/p$ 
for various values of $p$ and $N$ indeed
collapse when $t p^2/N$ is of order $1$.  On the other hand,
Fig.~\ref{fig:Nwk8resc}B indicates that the convergence towards
consensus is reached on a timescale of order $N^{a}$, with
$a \approx 1.4 \pm 0.1$, close to the
mean-field case $N^{3/2}$ and in strong contrast with the $N^3$
behavior of purely one-dimensional systems\footnote{
We also observe that the time to convergence scales as 
$p^{-1.4\pm.1}$; this is consistent with the fact that for 
$p$ of order $1/N$ one should recover an essentially one-dimensional
behaviour with convergence times of order $N^3$.
}. Moreover, as also observed
in mean-field~\cite{Baronchelli:2005}, the transition to the final 
consensus becomes more and more abrupt as the system size increases.

\begin{figure}
\centerline{
\includegraphics*[width=0.4\textwidth]{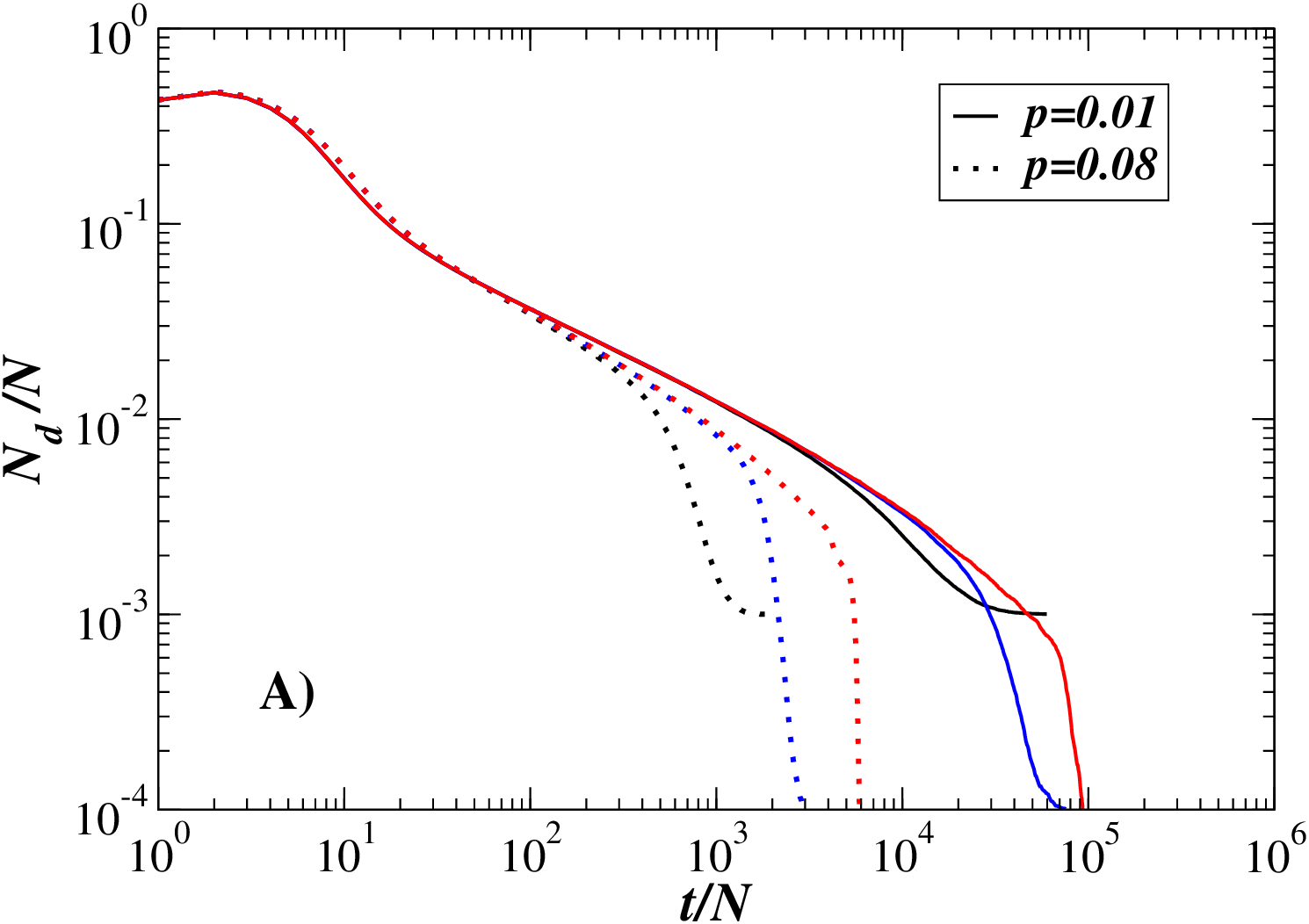}
\includegraphics*[width=0.4\textwidth]{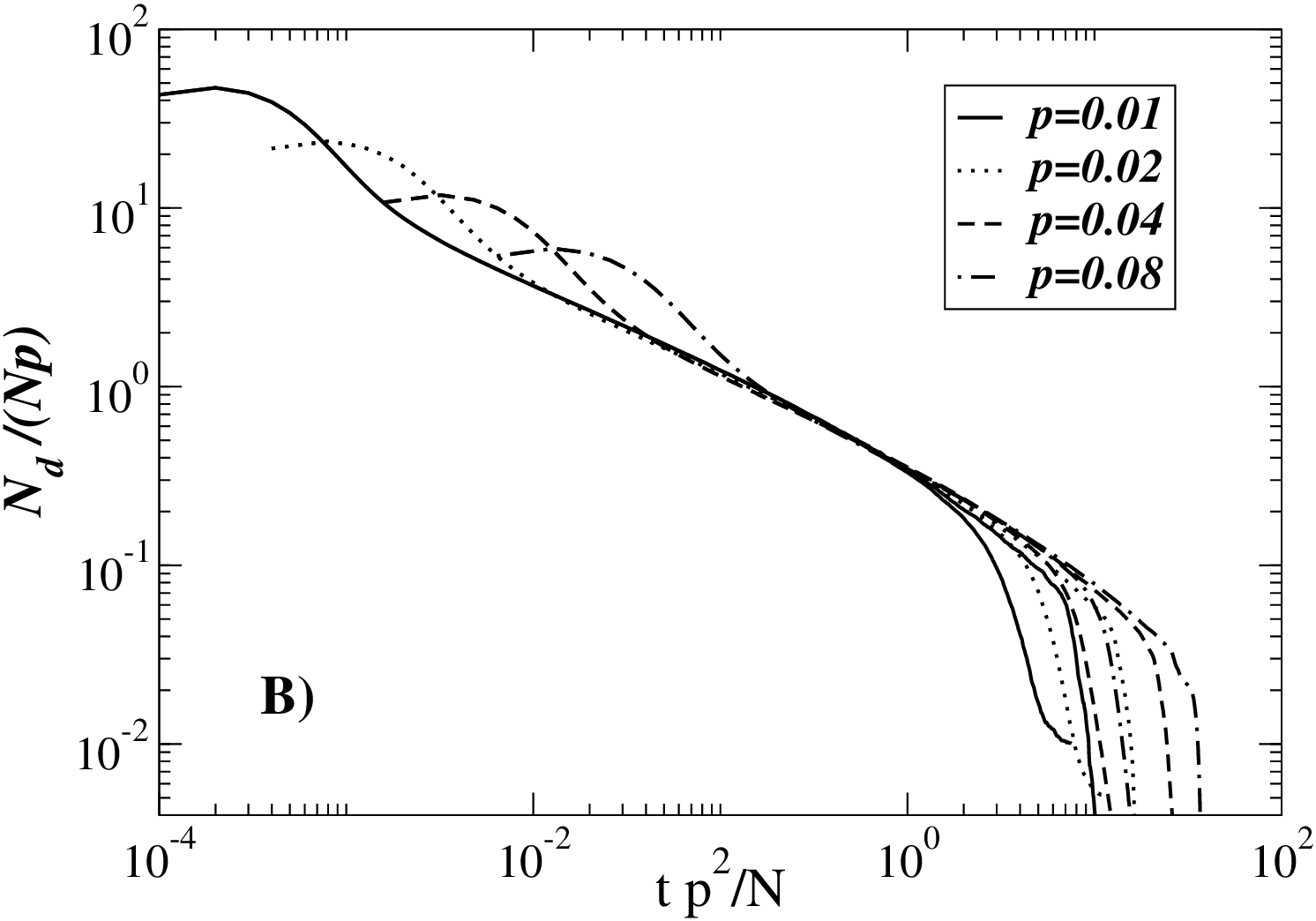}
}
\caption{A) (Color online) 
Number of different words in the system as a function of
time for $\langle k\rangle=8$, $p=10^{-2}$ and $p=8. 10^{-2}$ and
increasing sizes (from left to right): 
$N=10^3$ (black), $N=10^4$ (blue), $N=10^5$ (red).  B) Same data rescaled
in order to collapse the curves around the crossover time $N/p^2$.  As
in Fig.~\ref{fig:Nwk8resc}A), two values of the system size ($N=10^4$
and $N=10^5$) are displayed for each $p$.}
\label{fig:Ndk8}
\end{figure}

While the observation of Figs.~\ref{fig:Nwk8}B and~\ref{fig:Nwk8resc}B
could convey the impression that, after the memory peak, the system
tends to reach a stationary state whose length increases with $N$, the
analysis of the evolution of the number of {\em distinct} words
instead displays a continuous decrease (see Fig.~\ref{fig:Ndk8}):
during this apparent plateau therefore, the system is still evolving
continuously towards consensus by elimination of redundant words.
Figure~\ref{fig:Ndk8}A points out that this decrease is longer for
larger system sizes, while Fig.~\ref{fig:Ndk8}B shows that curves for
various system sizes and values of $p$ collapse when correctly
rescaled around the crossover time $N/p^2$.

\begin{figure}
\centerline{
\includegraphics*[width=0.4\textwidth]{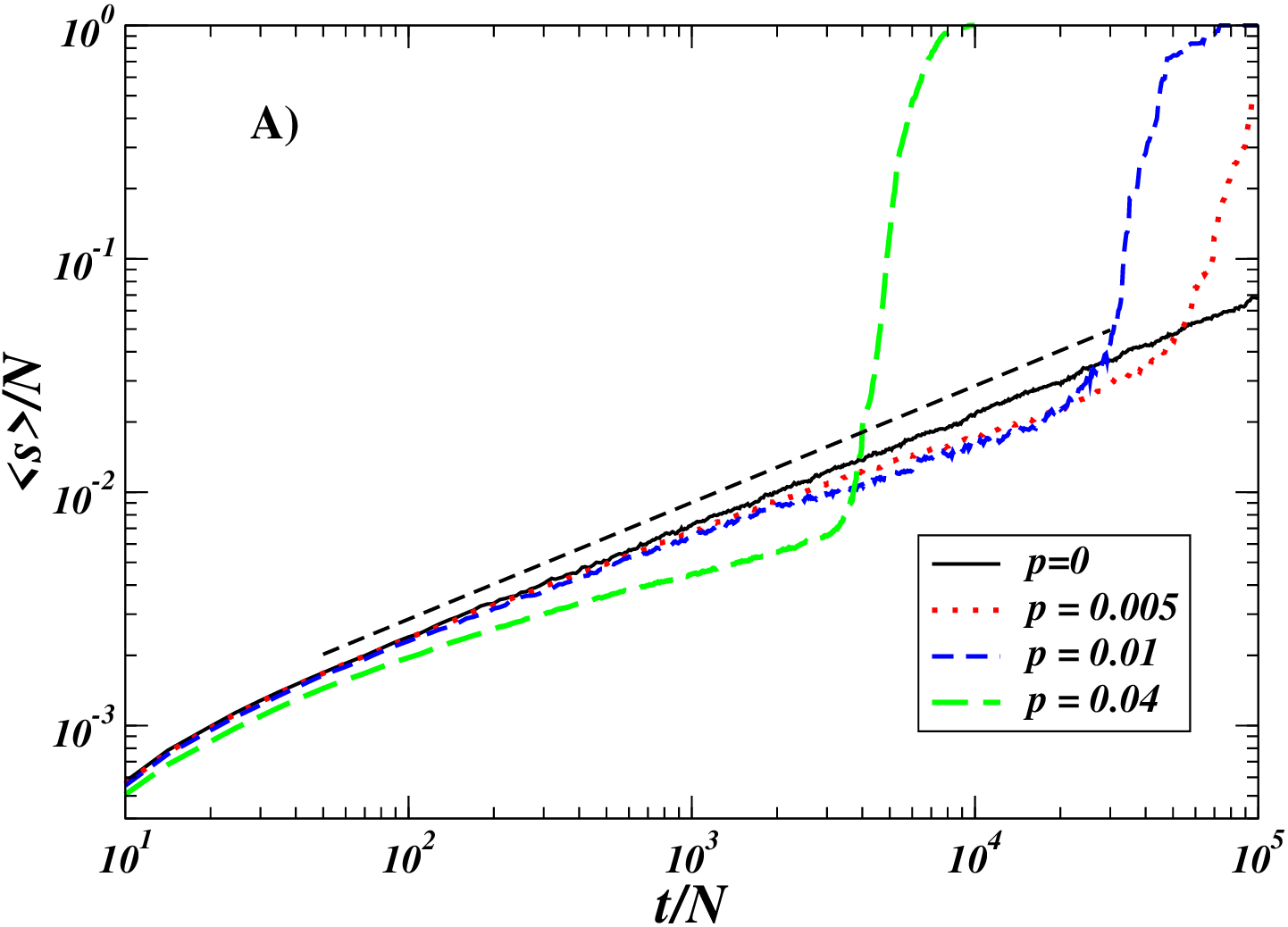}
\includegraphics*[width=0.4\textwidth]{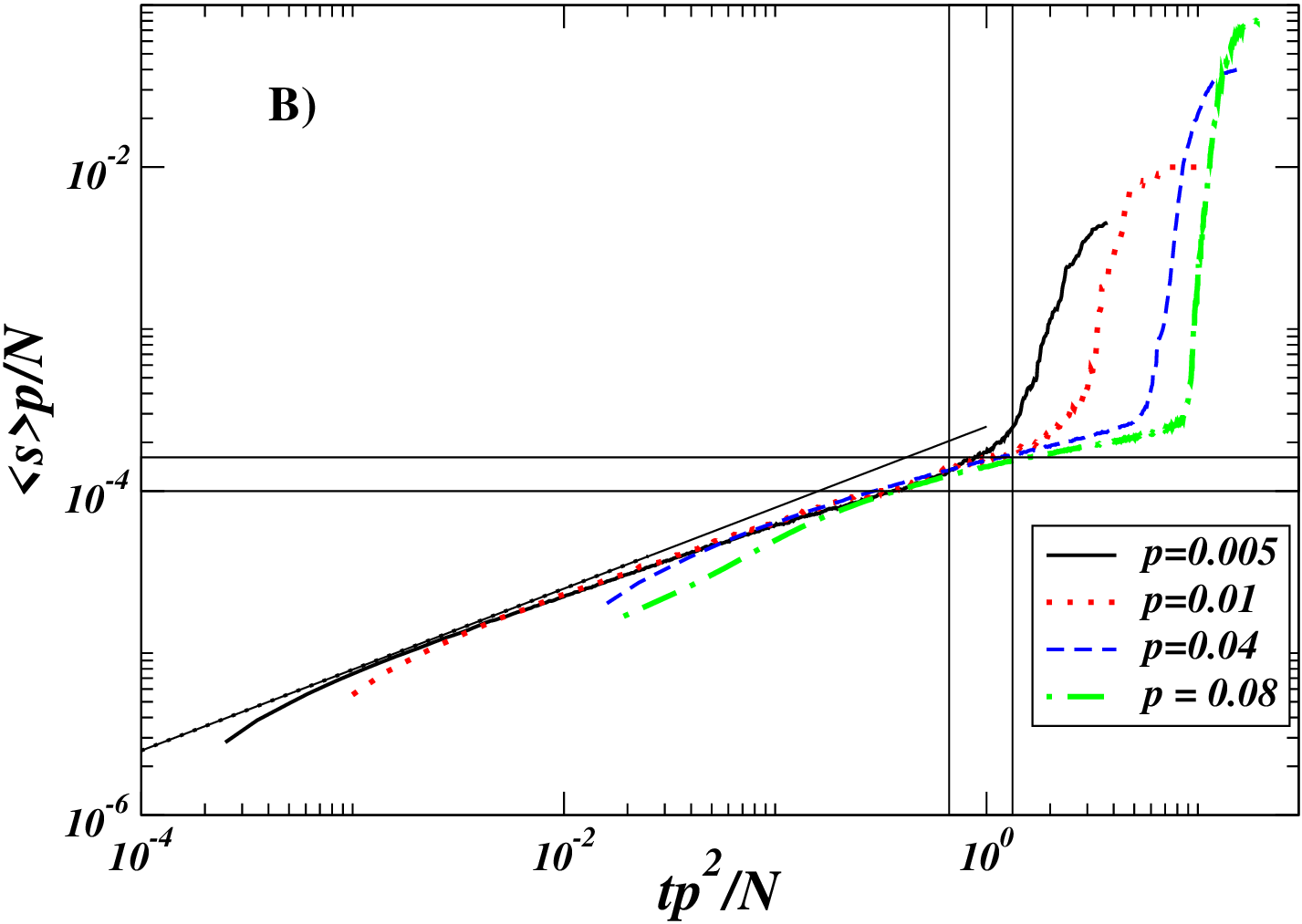}
}
\caption{(Color online) 
A) Evolution of the cluster size for $N=10^4$, various values of
$p$. At increasing $p$ the curves depart more and more from the
$t^{1/2}$ behavior through a slowing down of the cluster growth,
which however leads to a faster convergence.  B) Same curves rescaled 
around the crossover region.}
\label{fig:cluster}
\end{figure}

The combination of the results concerning average used memory and
number of distinct words correspond to a picture in which clusters of
agents sharing a common unique word compete during the time lapse
between the peak and the final consensus. It is thus interesting to
measure how the average cluster size evolves with time and how it
depends on the rewiring probability $p$. Figure~\ref{fig:cluster}
allows to compare the cluster size $\langle s \rangle$ evolution for
the one-dimensional case and for finite $p$. At $p=0$, a pure
coarsening law $\langle s \rangle \propto \sqrt{t}$ is observed.  As
$p$ increases, deviations are observed when time reaches the crossover
$p^2/N$, at a cluster size $1/p$, as was expected from the intuitive
picture previously developed (Fig.~\ref{fig:cluster} shows
the collapse of the curves of $\langle s \rangle p$ vs. $tp^2/N$
for $tp^2/N$ of order $1$).

Interestingly, the first deviation from the $\sqrt{t}$ law corresponds
to a {\em slowing down} of the cluster growth, correspondingly with
the slowing down observed in Fig.~\ref{fig:Nwk8}A.  Because of long-range
links, indeed, the clusters are locally more stable, due to the 
presence of an effective 'pinning' of interfaces near a shortcut. This 
effect is reminiscent of what happens for the Ising model on small-world 
networks~\cite{boyer} where, at low temperature, the local field 
transmitted by the shortcuts 
delay the passage of interfaces. Unlike Ising's zero 
temperature limit, however, the present dynamics only slows down and is
never blocked into disordered configurations.   

Strikingly, the final abrupt jump towards a unique cluster of size $N$
starts earlier and from smaller average cluster size as $p$ is
increased. Although not intuitive, this behavior can be explained as
follows. As $p$ increases, these clusters
are smaller and separated by more and more sites which have more than
one word in memory (hence a larger value of $N_w/N$ as $p$ increases)
and are more and more correlated. The sudden convergence to 
global consensus is thus obtained through a final fast agreement 
process between these sites.

\section{Conclusion and perspectives}

In summary this paper has explored how a non trivial interaction
topology could affect the dynamical approach to the emergence of
shared conventions in populations of agents. We have shown how
starting from a trivial one-dimensional like structure, the addition
of a finite number of long-range links ($p \gg 1/N$) leads to a
strong change in the dynamics, namely a drastic reduction of the
memory required to the agents and a significant acceleration of the
convergence process which passes from a $N^3$ dependence, typical of
the one-dimensional case to a $N^{1.4}$ scaling, close to the
mean-field one.  The overall dynamics occurs in two stages separated
by a crossover time scaling as $N/p^2$. These results open in our view
the way to a systematic investigation of this semiotic dynamics on
different classes of complex networks where the heterogeneity of the
nodes and the correlation between node degrees could play a major
role~\cite{inprep}.

\acknowledgements The authors thank E. Caglioti, M. Felici and
L. Steels for many enlightening discussions.  A. Baronchelli and
V.L. are partially supported by the EU under contract IST-1940
ECAgents project funded by the Future and Emerging Technologies
program (IST-FET) of the European Commission under the EU RD contract
IST-1940.  The information provided is the sole responsibility of the
authors and does not reflect the Commission's opinion. The Commission
is not responsible for any use that may be made of data appearing in
this publication.  A. Barrat and L.D. are partially supported by the
EU under contract 001907 ``Dynamically Evolving, Large Scale
Information Systems'' (DELIS).

\end{document}